\documentclass[usenatbib,referee]{ctextemp_mn2e}
\usepackage{epsfig}

\title[The formation of contact binaries]
{The detached-binary channel for the formation of contact binaries}
\author[D.Jiang et al.]{Dengkai Jiang$^{1,2}$\thanks{E-mail:
dengkai@ynao.ac.cn}, Zhanwen Han$^{1,2}$\thanks{E-mail:
zhanwenhan@ynao.ac.cn}, and Lifang Li$^{1,2}$ \\
$^{1}$Yunnan Observatories, Chinese Academy of Sciences, P.O. Box
110, Kunming, Yunnan Province, 650011, China\\
$^{2}$Key Laboratory for the Structure and Evolution of Celestial
Objects, Chinese Academy of Sciences, P.O. Box 110,\\
\ \ \ \  \ \ \ \ \ \ \ \ \ \ \ \ \ \ \ \ \ \ \ \ \ \ \ \ \ \ \ \ \ \
\ \ \  Kunming, Yunnan Province, 650011, China}
\begin{document}
\input ctextemp_psfig.sty
\date{Accepted .... Received .....; in original form ....}

\pagerange{\pageref{firstpage}--\pageref{lastpage}} \pubyear{2008}

\maketitle

\label{firstpage}

\begin{abstract}
The detached-binary channel is an important channel for the
formation of contact binaries, according to which a detached binary
might evolve into contact by evolutionary expansion of the
components, or angular momentum loss through the effect of magnetic
braking (MB). We have carried out a detailed binary population
synthesis (BPS) study of this channel, and obtained the parameter
regions for detached binaries to evolve into contact. Combining the
observations from the $\emph{Kepler}$ satellite with our results, we
found that the ratio of the birth rate of the progenitors of contact
binaries to that of contact binaries is greater than about 1.2. This
suggests that for the detached-binary channel, the progenitors can
be sufficient to produce observed contact binaries. In this channel,
we find that the distribution of orbital period for contact binaries
has a peak at about 0.25 days and a tail extending to longer
periods, and the formation timescale of contact binaries has a large
range from $\sim1$\,Myr to 15\,Gyr. These results show that the
detached-binary channel could explain satisfactorily the main
observational characteristics of contact binaries, in particular the
distribution of orbital period shown by the $\emph{Kepler}$
observations and the existence of very young contact binaries.
\end{abstract}\

\begin{keywords}
stars: activity -- binaries: close -- stars: formation -- stars:
evolution
\end{keywords}

\section{Introduction}
Contact binaries are interacting binaries in which two components
are overflowing their own Roche lobe and share a common envelope. In
general, contact binaries show periods from $\sim 0.2$ to 1.5\,d
\citep{Geske 2006, Paczynski 2006}. They are located near or just
above the main sequence (MS) \citep{Bilir 2005} and have the
shortest periods possible for binaries consisting of non-degenerate,
MS stars \citep{Baliunas 1985}. Mass transfer and energy transfer
between two components of contact binary would make the evolution of
the components different from that of single stars \citep{Yakut
2005, Jiang 2009}.

Contact binaries form an important class of binaries in several
respects. First, they could be used to investigate the merger
process of binaries. The recent observation of the remarkable system
V1309 Sco gave a direct evidence, for the first time, that contact
binaries indeed merge into the single objects \citep{Tylenda 2011}.
Secondly, contact binaries could be used to study the Galactic
structure because they have a high spatial frequency of occurrence,
are easy to detect and provide an absolute magnitude calibration
\citep{Rucinski 1997}. Finally, contact binaries play an important
role in stellar evolution as they are possible progenitors for some
objects, such as blue stragglers \citep{Eggen 1989, Mateo 1990,
Andronov 2006, Chen 2008}, FK Comae type stars \citep{Webbink
1976a}, $\lambda$ Bootis type stars \citep{Andrievsky 1997} and
Oe/Be stars \citep{Eggleton 2010, de Mink 2012, Jiang 2012b}.
Understanding the formation and evolution of contact binaries can
help to improve our understanding of the formation of these objects.

The detached-binary channel is an important channel for the
formation of contact binaries \citep{Rucinski 1986}, according to
which a detached binary might evolve into Roche lobe overflow
(RLOF), and subsequently into contact, by evolutionary expansion of
the components \citep{Webbink 1976b}, or by angular momentum loss
through the effect of MB \citep{Vilhu 1982}. From previous
observations, evidence was found that chromospherically active
binaries, which are one class of detached binaries, lose angular
momentum and evolve towards shorter orbital periods \citep{Demircan
1999, Karatas 2004, Eker 2006}. These binaries might be primary
candidates for the progenitors of contact binaries \citep{Eker
2006}. In addition, the presence of contact binaries in intermediate
age and old open clusters implies that they have evolved into
contact from detached progenitors \citep{Baliunas 1985, Rucinski
1998, Rucinski 2000, de Marchi 2007, Liu 2011}. Therefore, the
detached-binary channel is considered and investigated as the main
formation channel of contact binaries \citep{Rucinski 1986, Li
2007}.

However, based on the All-sky Automated Survey (ASAS) data,
\citet{Paczynski 2006} found that the number of detached binaries
with periods $P<1$\,d, which are believed to be the progenitors of
contact binaries, is insufficient to produce the number of observed
contact binaries. Therefore, they suggested that some contact
binaries might be formed in triple systems, where the inner binaries
with longer orbital period decrease the orbital period by Kozai
cycles and tidal friction, and evolve into contact binaries
\citep{Eggleton 2001}. In addition, \citet{Bilir 2005} found a small
group of very young ($<0.5$\,Gyr) contact binaries. They suggested
that the very young age of this group does not leave enough time for
detached binaries to evolve into contact, and these contact binaries
might be formed at the beginning of the main sequence or during the
pre-main-sequence contraction phase by a fission process
\citep{Roxburgh 1966}. \citet{van Eyken 2011} found that two contact
binaries are candidate members of 25 Ori or Orion OB1a association,
and they suggested that the $7-10$\,Myr age of the 25 Ori region is
too short for the formation of contact binaries from detached
binaries. Therefore, the formation of contact binaries seems to be
still an open question.

The detached-binary channel of contact binaries has been
investigated by many authors, e.g. \citet{Huang 1966}, \citet{Mestel
1968}, \citet{van't Veer 1979}, \citet{Vilhu 1982}, \citet{Maceroni
1991}, \citet{Stepien 1995, Stepien 2011}, \citet{Demircan 1999},
\citet{Tutukov 2004}, \citet{Bilir 2005} and \citet{Eker 2006}.
\citet{Vilhu 1982} calculated the period evolution of detached
binaries with a total mass 2\,M$_{\rm \odot}$ that evolve into
contact by considering the angular momentum loss. He suggested that
contact binaries could be produced from detached binaries in old
($\sim 5 \times 10^9$yr) and in intermediate age ($\sim 5 \times
10^8$yr) clusters, but in very young clusters only if the initial
period is sufficiently short. Therefore, the formation of contact
binaries from detached binaries depends on the initial distribution
of parameters of detached binaries \citep{Vilhu 1982, Eker 2008}.
Moreover, the rate of angular momentum loss through the effect of MB
is another important but rather poorly known factor \citep{Vilhu
1982}. The evolution of orbital period determines which detached
binary can evolve into contact, and is very different for different
models of MB \citep{Vilhu 1982, Bradstreet 1994, Stepien 1995}.
Therefore, further investigation is needed of the detached-binary
channel for the formation of contact binaries.

The purpose of this paper is to investigate the formation of contact
binary from detached binary by using binary population synthesis.
The outline of this paper is as follows. In Section 2, we describe
the BPS method. The results are shown in Section 3. Finally, we give
the discussion and conclusions in Section 4.

\section{binary population synthesis}
In the detached-binary channel, the primary of detached binary would
first fill its Roche lobe and transfer some of its mass to the
secondary. If the secondary also fills its Roche lobe in response to
thermal time-scale or nuclear time-scale mass transfer, these
binaries would evolve into contact when both components are still MS
stars. To determine whether the binaries evolve into contact, we use
Hurley's rapid binary evolution code \citep{Hurley 2000, Hurley
2002}, and perform seven sets of simulations (see Table 1) in the
BPS study. In each simulation, we follow the evolution of $10^6$
sample binaries (Z $=0.02$) from the star formation to the formation
of contact binaries. If both components of a binary are MS stars and
fill their Roche lobes, we assume this results in a contact binary,
and the properties of the contact binary at the moment of their
formation are obtained. In addition, in order to understand the
detached-binary channel better, it is more instructive to compare
the theoretical distribution of orbital periods and temperatures
with that of the observed binaries. \citet{Prsa 2011} compiled a
Eclipsing Binary Catalogue based on the observation of the
$\emph{Kepler}$ space mission. This catalogue is updated by
\citet{Slawson 2011} and \citet{Matijevic 2012}, and an online
version is maintained at http://keplerEBs.villanova.edu and on
Mikulski Archive for Space Telescopes (MAST),
http://archive.stsci.edu/kepler. We will compare our results with
the observed binaries in this catalogue.

\subsection{Monte Carlo simulation parameters}
In the BPS study, the Monte Carlo simulations require the following
physical input: the initial mass function (IMF) of the primaries,
the initial mass-ratio distribution, the distribution of initial
orbital separations, the eccentricity distribution of binary orbit,
and the star formation rate (SFR) \citep[e.g.][]{Han 2002, Han 2003,
Liu 2009, Wang 2010}:
\begin{itemize}
\item[(i)]
We use a simple approximation to the IMF of \citet{Miller 1979} and
the initial mass of the primary ($M_{10}$) is generated using a
formula of \citet{Eggleton 1989},
\begin{equation}
M_{10} =\frac{0.19X}{(1-X)^{0.75}+0.032(1-X)^{0.25}},
\end{equation}
where $X$ is a random number uniformly distributed between 0 and 1.
The study of IMF by \citet{Kroupa 1993} supports this IMF.

\item[(ii)]
We take a constant mass-ratio ($q_{0}$) distribution (set 1, 3-7),
\begin{equation}
n(q_{0}) =1, \;\;\;\;  0 \leq q_{0} \leq 1
\end{equation}
where $q_0=M_{20}/M_{10}$, and $q_0, M_{20}$ are the initial mass
ratio and the initial mass of the secondary \citep{Mazeh 1992,
Goldberg 1994}. In order to study the influence of the mass ratio
distribution, we also take an alternative mass-ratio distribution
where both components are chosen randomly and independently from the
same IMF (set 2).

\item[(iii)]
We assume that all stars are members of binary systems and that the
distribution of separations is constant in log\,$a$ for wide
binaries, where $a$ is the orbital separation and falls of smoothly
at small separation
\begin{equation}
an(a)=\left\{ \begin{array}{ll}
                    \alpha_{\rm sep} (\frac{a}{a_{0}})^m,  & \;\;\;\;\;\;\;\;\;\; \mbox{ $a \leq a_{0}$,} \\
            \alpha_{\rm sep},  & \;\;\;\;\;\; \mbox{ $a_{0} < a < a_{1} $,}
           \end{array}
       \right.
\end{equation}
where $\alpha_{\rm sep}\approx 0.070$, $a_0=10$ R$_{\rm \odot}$,
$a_1=5.75 \times 10^6$ R$_{\rm \odot}= 0.13$ pc, and $m \approx
1.2$. This distribution implies that the numbers of wide binary
systems per logarithmic interval are equal, and that about 50\% of
stellar systems have orbital periods less than 100 yr
\citep[][HPE95]{Han 1995}. To investigate the effect of initial
distribution of orbital separation, we also take a well-determined
period distribution of solar-type MS binaries
\citep[][DM91,R10]{Duquennoy 1991,Raghavan 2010}, and the orbital
period ($P_0$) is generated using a formula of \citet{Eggleton
2006},
\begin{equation}
P_{0} =5.0 \times 10^4 (\frac{X}{1-X})^{3.3},
\end{equation}
where $X$ is a second, independent, random number uniformly
distributed between 0 and 1. This distribution is a good fit to the
Duquennoy-Mayor distribution \citep{Eggleton 2006}.

\item[(iv)]
A circular orbit is assumed for all binaries.

\item[(v)]
We assume either a single starburst or a constant SFR over the
15Gyr. For the case of a single starburst, we assume a burst
producing 10$^6$ binaries to investigate the formation of contact
binaries in a star cluster. In the case of a constant SFR, we assume
SFR$=5\,$M$_{\rm \odot}\,$yr$^{-1}$ in a similar way to the study of
supernova rate \citep{Wang 2010}.

\end{itemize}

\begin{table}
\begin{minipage}{85mm}
\caption{Sets in different simulations (metallicity Z $=0.02$).
$n(q_0)=$ initial mass ratio distribution; MB $=$ the Model of
magnetic braking; $q_{\rm conv}=$ the expression of the mass
fraction of the surface convective envelope ($q_{\rm conv}$) in MB
models; $a =$ the distribution of orbital separation.}
\begin{tabular}{cccccc}
\hline
Set & $n(q_0)$ & ${\rm MB}$ & $q_{\rm conv}$& $a$\\
\hline
$1$ & ${\rm Constant}$     & ${\rm HTP02}$      & ${\rm linear}$         & ${\rm HPE95}$ \\
$2$ & ${\rm Uncorrelated}$ & ${\rm HTP02}$      & ${\rm linear}$         & ${\rm HPE95}$ \\
$3$ & ${\rm Constant}$     & ${\rm HTP02}$      & ${\rm exponential}$    & ${\rm HPE95}$ \\
$4$ & ${\rm Constant}$     & ${\rm S95,S11}$    & ${\rm exponential}$    & ${\rm HPE95}$ \\
$5$ & ${\rm Constant}$     & ${\rm SPT00}$      & ${\rm exponential}$    & ${\rm HPE95}$ \\
$6$ & ${\rm Constant}$     & ${\rm no \; MB}$   & $-$                    & ${\rm HPE95}$ \\
$7$ & ${\rm Constant}$     & ${\rm HTP02}$      & ${\rm exponential}$    & ${\rm DM91,R10}$ \\
\hline
\end{tabular}
\end{minipage}
\end{table}

\subsection{The magnetic braking}
The rate of angular momentum loss (AML) by magnetic braking (MB) is
a very important parameter for the orbital evolution of detached
binaries, and therefore for the formation of contact binaries from
detached binaries. In simulation set 1, we adopt the AML rate by MB
used by \citet[][HTP02]{Hurley 2002}, which is expressed as
\begin{equation}
\frac{{\rm d}J_{\rm MB}}{{\rm d}t} =-5.83\times 10^{-16}q_{\rm
conv}(R\Omega_{\rm spin})^3{\rm M_\odot}{\rm R_\odot}^2{\rm
yr}^{-2},
\end{equation}
where $q_{\rm conv}$, $R$ and $\Omega_{\rm spin}$ are the mass
fraction of the surface convective envelope ($M_{\rm env}/M$), the
radius of the component in solar units, and the spin frequency of
the component in units of year$^{-1}$.

In order to investigate the effect of the magnetic braking, we also
take other descriptions of the AML by MB. In set 4, we adopt the
description of AML by MB from \citet[][S95,S11]{Stepien 1995,
Stepien 2011}:
\begin{equation}
\frac{{\rm d}J_{\rm MB}}{{\rm d}t} =-4.9\times
10^{41}(R_1^2M_1+R_2^2M_2)/P,
\end{equation}
where, $M_{1,2}$ and $R_{1,2}$ are the masses and radii for the
primary and the secondary in solar units, $P$ is the period of
binary in days, $J_{\rm MB}$ is in cgs units and time is in years.
This description of MB was derived and calibrated from observations
of spin-down of single stars \citep{Stepien 1995}, and we assume
that the binary systems are in synchronous rotation. In addition,
the expression for the AML by MB from \citet[][SPT00]{Sills 2000} is
used in set 5:
\begin{equation}
\frac{{\rm d}J_{\rm{MB}}}{{\rm d}t}=\left\{
 \begin{array}{lc}
-K \omega^3 \left(\frac{R}{R_{\odot}}\right)^{0.5}
\left(\frac{M}{M_{\odot}}\right)^{-0.5},  \hspace{20pt} \omega \leq \omega_{\rm{crit}},\\
-K \omega_{\rm{crit}}^2 \omega
\left(\frac{R}{R_{\odot}}\right)^{0.5}
\left(\frac{M}{M_{\odot}}\right)^{-0.5},  \ \omega >
\omega_{\rm{crit}},\\
\end{array}\right.
\end{equation}
where $K=2.7\times10^{47}$ g\,cm$^2$\,s \citep{Andronov 2003},
$\omega$ is the angular velocity of synchronized binary, and $M, R$
and $\omega_{\rm{crit}}$ are the mass of component, the radius of
component and the critical angular velocity at which the angular
momentum loss rate reaches a saturated state \citep{Krishnamurthi
1997}. This description of the AML by MB adopt a modified
\citet{Kawaler 1988} AML rate with $N=1.5$ wind law \citep{Sills
2000}.

The effect of MB is expected to be reduced when the convective
envelope becomes too small \citep{Podsiadlowski 2002}, and this is
not considered in Equations (6) and (7). Therefore, we add an ad hoc
factor ${\rm exp}(-0.02/q_{\rm{conv}}+1) \; {\rm if}\;
q_{\rm{conv}}<0.02$ in these equations following the suggestion of
\citet{Podsiadlowski 2002}. On the other hand, to investigate the
effect of the dependence of MB on $q_{\rm conv}$, in set 3 we retain
the functional dependence of the braking on stellar radius and spin
frequency given by Equation (5), but use the ad hoc factor ${\rm
exp}(-0.02/q_{\rm{conv}}+1)$ instead of $q_{\rm conv}$ in set 3. For
the exponential form in sets 3-5 and 7, the effect of MB is a
decrease for stars with mass greater than $\sim 1.0$\,M$_{\rm
\odot}$ (corresponding to $q_{\rm conv}$ $<$0.02), while the effect
of MB does not depend on $q_{\rm conv}$ for star with mass smaller
than this mass. For the linear form in sets 1 and 2, the effect of
MB decreases with decreasing $q_{\rm conv}$, when the mass of star
increases from 0.35 to 1.25\,M$_{\rm \odot}$.

\section{results}

%\subsection{The influence of MB models}
%\subsection{Formation of contact binaries}
\subsection{The orbital evolution for the formation of contact binaries}

\begin{figure*}
\centerline{\psfig{figure=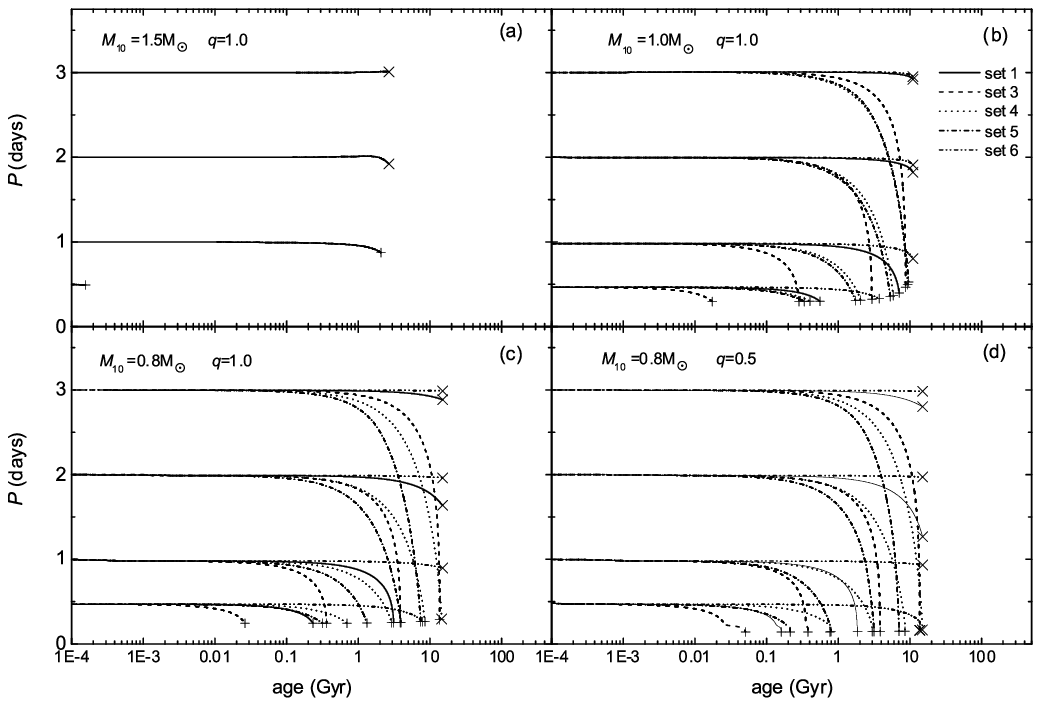,width=15cm}} \caption{Orbital
evolution of typical detached binaries in simulation sets with
different MB models. Solid curves are for Hurley's MB model (set 1).
Dashed curves are for Hurley's MB model with another expression of
the mass fraction of the convective envelope (set 3). Dotted and
dash-dot curves are for the MB models given by \citet{Stepien 1995,
Stepien 2011} (set 4) and \citet{Sills 2000} (set 5). Dash-dot-dot
curves are for no MB (set 6). Pluses indicate the position where the
system evolves into contact. Crosses represent systems that can not
evolve into contact while both components are still MS stars. }
\label{fig2}
\end{figure*}

The orbital evolution of detached binaries is very important for the
formation of contact binaries in the detached-binary channel. In
Fig. 1 we present the evolution of orbital periods of the typical
examples in the simulations with different MB models. We do not show
the orbital evolution of binary systems in the simulation sets 2 and
7, which are the same as that in the simulation sets 1 and 3. For
systems with $M_{10}=1.5$\,M$_{\rm \odot}$ and $q=1.0$ (Fig. 1a),
the orbital periods are almost constant. This is because there is no
MB effect for the components with mass larger than 1.25\,M$_{\rm
\odot}$ that do not have a convective envelope \citep{Hurley 2000}.
Binary systems with very short period ($P_0=0.5, 1.0$\,d) can evolve
into contact by evolutionary expansion of the components while both
components are still MS stars. For systems with period $P_0=2.0,
3.0$\,d, the components do not fill their Roche lobes, and do not
evolve into contact before they leave MS.

For systems with the mass of components less than 1.25\,M$_{\rm
\odot}$, such as systems with $M_{10}=1.0, 0.8$\,M$_{\rm \odot}$
(Figs 1b, c and d), the orbital periods of systems with MB effect
(solid, dashed, dotted and dash-dot curves for set 1, 3, 4 and 5)
decrease more quickly than that with no MB effect (dash-dot-dot
curves for set 6). This results in the formation of contact binaries
for some systems with long period ($P_0 \sim$ 3\,d). However, the
results are different for systems with different MB Models. For
example, binary systems with $M_{10}=1.0$\,M$_{\rm \odot}$, $q_0=1$
and $P_0=2.0, 3.0$\,d can evolve into contact in set 3, 4 and 5, but
not in set 1 as shown in Fig. 1(b). In addition, even in the same MB
model, the evolution of the orbital period for systems with
$M_{10}=0.8$ (Fig. 1c) are different from that for systems with
$M_{10}=1.0$ (Fig. 1b). We show the evolution of binary systems in
Fig. 1(d) that have different mass ratio ($q_0=0.5$) from those in
Fig. 1(c). It is obvious that the mass ratio also affects the
orbital evolution, and therefore the formation of contact binaries
due to the dependence of the magnetic braking on stellar mass.

%\subsection{The distribution of the progenitors of contact binaries}
%\subsection{Distribution of initial parameters of detached binaries for the formation of contact binaries}
%\subsection{Progenitors of contact binaries}
\subsection{Initial parameters for the progenitors of contact binaries}

%\begin{figure*}
%\centerline{\psfig{figure=fig2.eps,width=12.5cm}}

\begin{figure}
%\epsscale{1.0}
%\plotone{fig2.eps}
\centerline{\epsfig{figure=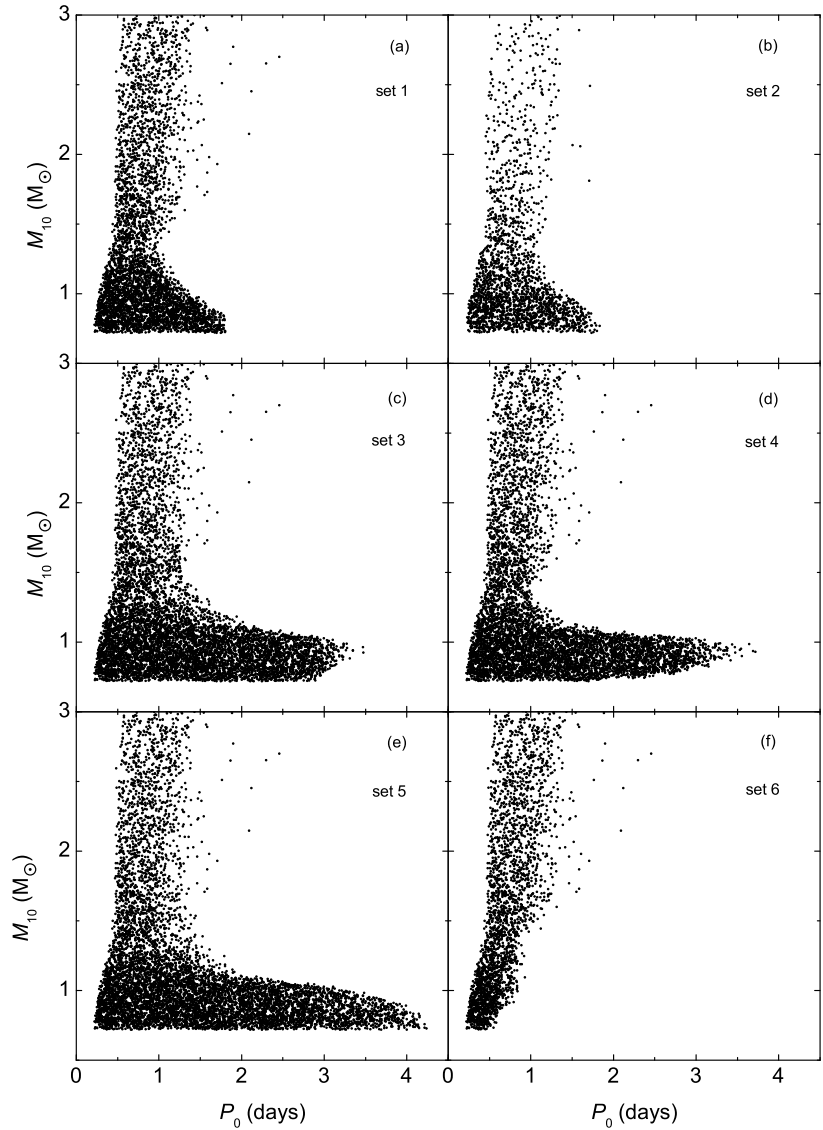,width=12.5cm}}

\caption{The initial distribution of detached binaries from the
simulation sets 1-6 in the $P_0-M_{10}$ plane that produce contact
binaries, where $P_0$ is the initial orbital period and $M_{10}$ is
the initial mass of the primary.} \label{fig2}
%\end{figure*}
\end{figure}

In Fig. 2, we present the initial distribution of detached binaries
that produce contact binaries in the $P_0-M_{10}$ plane. Fig. 2(a)
represents the initial parameters of the progenitors of contact
binaries in the simulation with the MB model of \citet{Hurley 2002}
(set 1). For systems with $M_{10}<0.7$\,M$_{\rm \odot}$, the primary
is deeply convective, and these systems experience unstable mass
transfer \citep{Hurley 2002}. Therefore, they can not form stable
contact binaries \citep{Jiang 2012}. It should be noted that this
mass limit for deeply convective stars is not a sharp limit but a
gradual transition \citep{Politano 1996}, and therefore this limit
for the formation of contact binaries is also not as sharp as
assumed, but depends on the mass ratio \citep{Jiang 2012}. The lower
limit of the initial orbital period (left boundary) for the
formation of contact is about $0.22-0.5$\,d, which is set by the
condition of initial orbital period that neither of the components
fills the Roche lobe at birth. The upper limit of the initial
orbital period (right boundary) depends on the initial mass of the
primaries, and is caused by the constraints that detached binaries
have to evolve into contact in 15Gyr and that both components are
still MS stars. The upper limit of the initial orbital period
decreases from 1.8\,d at $M_{10}\sim0.7$\,M$_{\rm \odot}$, to $\sim
1.0$\,d at $M_{10}\sim 1.3$\,M$_{\rm \odot}$ due to the decreasing
fraction of the convective envelope, which leads to a weaker MB. The
simulation with a mass-ratio distribution with uncorrelated binary
components (set 2) has a similar region to set 1 as shown in Fig.
2(b), although the number is smaller. This is because more binary
systems have very small mass ratios, and the secondaries merge into
the primaries following the onset of RLOF due to dynamical mass
transfer \citep{Hurley 2002}.

Fig. 2(c) shows the distribution of the formation of contact
binaries for the simulation (set 3) with different expression of the
mass fraction of the convective envelope in the MB model from set 1.
The main difference between this set and the previous one (set 1) is
that for systems with the mass of the primaries less than $\sim
1.3$\,M$_{\rm \odot}$, the upper limit of the initial orbital period
is about $3-3.5$\,d, and is much longer than that in set 1. We also
note that the upper limit of the initial orbital period in the
simulations with the MB model of \citet{Stepien 1995, Stepien 2011}
(set 4) and the MB model of \citet{Sills 2000} (set 5) are similar
to that in set 3, and are much longer than that in set 1 as shown in
Figs 2(d) and (e). The maximum value of the upper limit of the
initial orbital period is about 3.7\,d in set (4), and 4.2\,d in set
(5). This suggests that the expression for the dependence of MB on
$q_{\rm conv}$ has a great influence on the formation of contact
binary. We do not show the distribution of the formation of contact
binaries in set 7 that is similar to the distribution in set 3.

Fig. 2(f) shows the distribution of the formation of contact
binaries for the simulation with no MB (set 6). It is shown that for
systems with $M_{10}>1.3$\,M$_{\rm \odot}$, the distribution is
similar to that in other sets. However, for systems with
$M_{10}<1.3$\,M$_{\rm \odot}$, the upper limit of initial orbital
period is about $0.6-1$\,d, which is significantly shorter than that
in other sets. Therefore, only detached binaries with very short
orbital period could evolve into contact if systems have low mass
primaries. This suggests that the effect of MB is important for the
formation of contact binaries with low mass components.

\subsection{Progenitors of contact binaries: comparison with observations}

\begin{figure}
\centerline{\psfig{figure=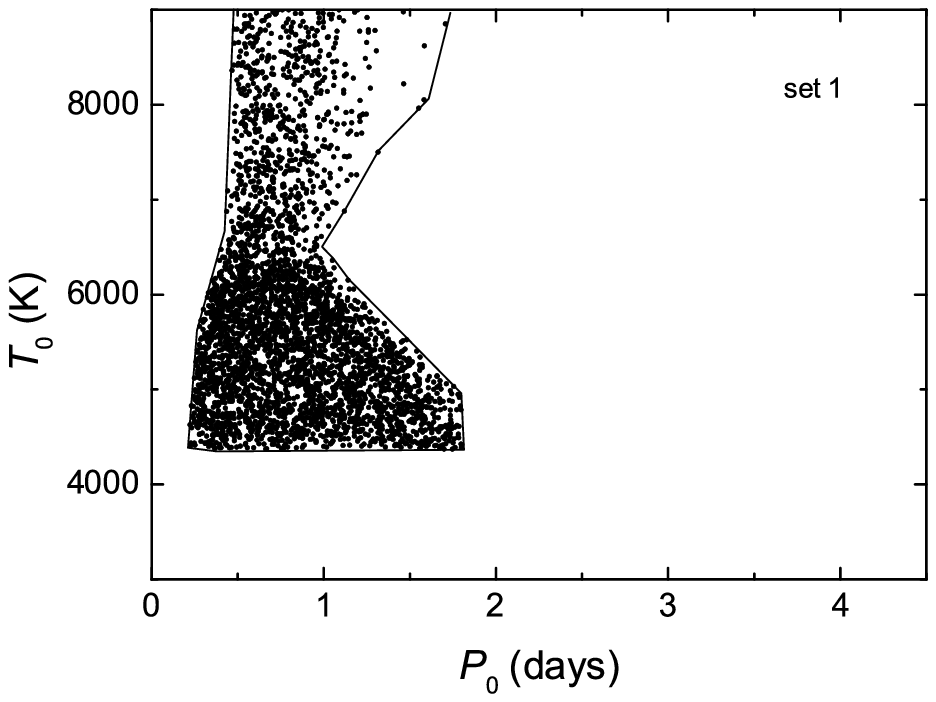,width=7.5cm}} \caption{The
distribution of the progenitors of contact binaries in the $P_0-T_0$
plane that produce contact binaries, where $P_0$ is the initial
orbital period and $T_0$ is the zero-age main-sequence effective
temperature of the primary. Solid curves mark the region for
detached binaries that evolve into contact.} \label{fig5}
\end{figure}

\begin{figure}
\centerline{\psfig{figure=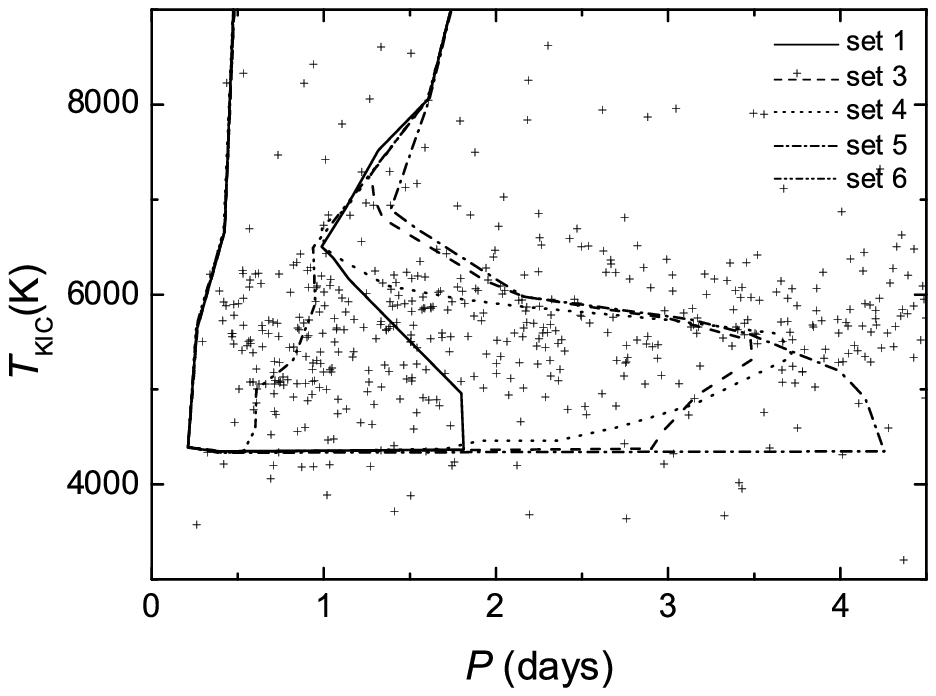,width=7.5cm}} \caption{The
distribution of the observed detached eclipsing binaries in the
Eclipsing Binary Catalogue \citep{Prsa 2011, Slawson 2011, Matijevic
2012}. Solid curves show the parameter region for detached binaries
that evolve into contact in set 1 as shown in Fig. 3, and dashed,
dotted, dash-dot and dash-dot-dot curves show the parameter regions
in simulation sets $3-6$.} \label{fig5}
\end{figure}

In order to understand the detached-binary channel better, we
compare our theoretical distribution of progenitors of contact
binaries with that of the observed detached binaries in the
Eclipsing Binary Catalogue given by \citet{Prsa 2011},
\citet{Slawson 2011} and \citet{Matijevic 2012}. Fig. 3 shows the
theoretical distribution of initial orbital period ($P_0$)$-$initial
temperature ($T_0$) for detached binaries in set 1 that evolve into
contact, where $T_0$ is taken as the zero-age main-sequence
effective temperature of the primary. Solid curves mark the
parameter region for detached binaries that evolve into contact. In
Fig. 4 we plot the distribution of observed detached binaries in the
Eclipsing Binary Catalogue \citep{Prsa 2011, Slawson 2011, Matijevic
2012}, where $T_{\rm KIC}$ is the $\emph{Kepler}$ Input Catalog
effective temperature. The curves show the regions for detached
binaries that evolve into contact in different simulations sets.

It is apparent that many observed detached binaries are located in
the formation region of contact binaries as shown in Fig. 4. For the
simulation with no MB (set 6), the number of detached binaries in
the formation region of contact binaries is 75. In the simulations
(set 1, 3, 4 and 5) considering the effect of MB, the numbers are
179, 312, 292 and 331, respectively. These detached binaries might
be the progenitors of contact binaries, and their number seems to be
comparable to the number of observed contact binaries, which is 469
in the Eclipsing Binary Catalogue \citep{Prsa 2011, Slawson 2011,
Matijevic 2012}. In addition, detached binaries located outside of
the formation region of contact binaries can not result in stable
contact binaries.

To estimate whether the progenitors of contact binaries are
sufficient to produce the number of observed contact binaries, we
roughly derive the actual ratio of the number of observed
progenitors of contact binaries to that of observed contact binaries
by taking into account observational selection effects. The
detection probability of contact binaries should be independent of
orbital period in first order approximation, because the size of the
Roche lobe is proportional to the orbital separation. Then, for
contact binaries at a given orbital period $P$, the observed number
$N_{\rm obs, con}(P)$ can be expressed as
\begin{equation}
N_{\rm obs, con}(P)={\rm b} \times N_{\rm 0, con}(P),
\end{equation}
where b is a (constant) detection probability for contact binaries,
and $N_{\rm 0, con}(P)$ is the actual number of contact binaries
with orbital period $P$. \citet{Maceroni 1999} suggested that the
probability of discovering an eclipsing system scales as the inverse
square of its orbital separation, $\propto a^{-2} \propto P^{-4/3}$.
Therefore, for detached progenitor binaries at a given orbital
period $P$, the observed number $N_{\rm obs, pro}(P)$ can be
obtained by
\begin{equation}
N_{\rm obs, pro}(P)={\rm c} \times P^{-\frac{4}{3}} \times N_{\rm 0,
pro}(P),
\end{equation}
where c is a constant, and $N_{\rm 0, pro}(P)$ is the actual number
of detached progenitor binaries with orbital period $P$. We assume
that $N_{\rm obs}(P) = 1$ for each binary in the Eclipsing Binary
Catalogue. Then, for the contact binary population and their
progenitor population of detached binaries, their actual numbers can
be expressed as
\begin{equation}
N_{\rm con}= \sum N_{\rm 0, con}(P_i) =\sum \frac{1}{{\rm b}}
\;\;\;\;\;\;\;\;\;\; i=1, {\rm n_{\rm con}},
\end{equation}
and
\begin{equation}
N_{\rm pro}= \sum N_{\rm 0, pro}(P_j) =\sum \frac{1}{{\rm c} \times
P_j^{-\frac{4}{3}}}  \;\;\;\;\;\;\;\;\;\; j=1, {\rm n_{\rm pro}},
\end{equation}
where ${\rm n_{\rm con}}$ and ${\rm n_{\rm pro}}$ are the observed
numbers of contact binaries and the progenitors of contact binaries
in the Eclipsing Binary Catalogue.

We can obtain a lower limit of the actual ratio of the number of
progenitors of contact binaries to that of contact binaries without
the values of b and c. The range of inclinations for which a contact
binary can be detected is larger than that for a detached binary
with same orbital period and stellar masses, if the orientations of
the orbital planes are assumed to be randomly distributed for
contact binaries and detached binaries. This is because both
components of contact binaries overfill their Roche lobes, while
both components of detached binaries are inside their Roche lobes.
Therefore, the detection probability for a contact binary should be
larger than that of a detached binary at any orbital period, in
other words: ${\rm b} > {\rm c} \times P^{-4/3}$ for any value of
$P$. Then, we obtain
\begin{equation}
N_{\rm con} =\sum \frac{1}{{\rm b}} < \sum \frac{1}{{\rm c} \times
P_i^{-\frac{4}{3}}} \;\;\;\;\;\;\;\;\;\; i=1, {\rm n_{\rm con}}.
\end{equation}
Finally, we can get a lower limit of the ratio $N_{\rm pro}/N_{\rm
con}$, and for simulation set 3, this ratio is larger than about
2.5.

To compare the birth rates of the progenitor population with that of
the contact binary population, the relative lifetimes in the
detached and contact phases also need to be considered. For the
detached binaries that evolve into contact in the simulation set 3,
the mean lifetime of detached phase is about 2.1\,Gyr. \citet{Jiang
2012a} show that the lifetime of contact phase is about 4\%-10\% of
the main sequence lifetime of the primaries. We adopt the middle
value 7\%, and get the lifetimes of contact phase for every contact
binaries in simulation set 3. The mean lifetime of contact phase is
about 1.04\,Gyr for these contact binaries . Therefore, the ratio of
the lifetime of detached phase to that of contact phase is $\approx
2$. Combining the lower limit to the ratio $N_{\rm pro}/N_{\rm
con}$, the ratio of the birth rate of the progenitors of contact
binaries to that of contact binaries is estimated roughly to be
greater than about 1.2.

\subsection{The distribution of orbital periods of contact binaries}

\begin{figure}
\centerline{\psfig{figure=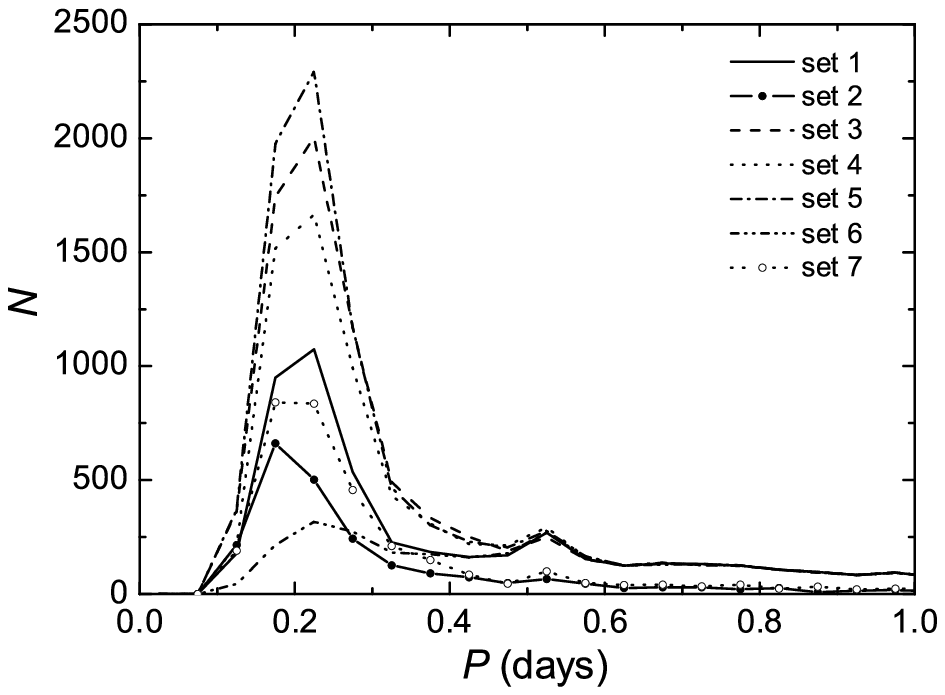,width=7.5cm}} \caption{The
distribution of orbital periods of contact binaries at the moment of
the formation from all the simulation sets.} \label{fig3}
\end{figure}

\begin{figure}
\centerline{\psfig{figure=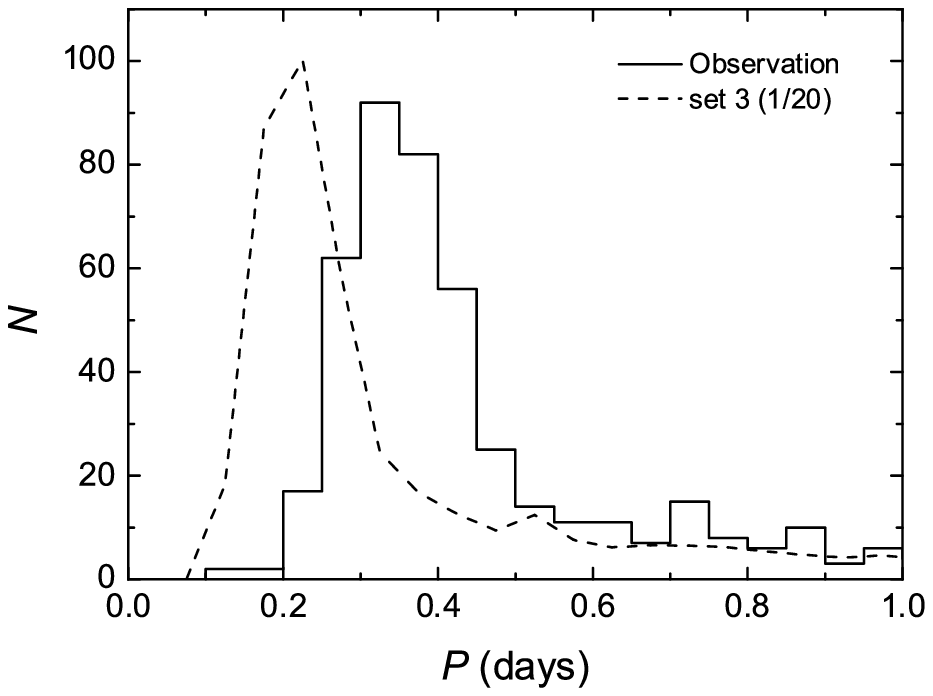,width=7.5cm}} \caption{The
distribution of orbital period of contact binaries from the
detached-binary channel in simulation set 3, which has been rescaled
by a factor of 1/20 for clarity. The solid histogram represents the
observed contact binaries in the Eclipsing Binary Catalogue given by
\citet{Prsa 2011}, \citet{Slawson 2011} and \citet{Matijevic 2012}}
\label{fig3}
\end{figure}

Fig. 5 shows the distribution of orbital period of contact binaries
at the moment of the formation for simulation sets $1-7$. The
distribution of contact binaries has a peak around 0.25\,d and a
long tail extending beyond 1\,d. This peak comes from contact
binaries mainly formed by the effect of MB, although the formation
of these contact binaries might be also affected by evolutionary
expansion of the components. The number of contact binaries in this
peak strongly depends on the models of MB, the dependence of MB on
$q_{\rm conv}$, the initial mass-ratio distribution and the initial
orbital period distribution. In Fig. 6 we compare the distribution
of the orbital period in simulation set 3 with that of observed
contact binaries in the Eclipsing Binary Catalogue \citep{Prsa 2011,
Slawson 2011, Matijevic 2012}. The distribution of observed contact
binaries also show a peak and a tail extending to longer periods.
This is similar to the theoretical distribution in the
detached-binary channel. However, it should be noted that the
distribution of observed contact binaries has a translation to
longer period, and its peak is located at $P\sim0.35$\,d.

\subsection{The distribution of contact binaries in $P-T$ diagram}

\begin{figure}
\centerline{\psfig{figure=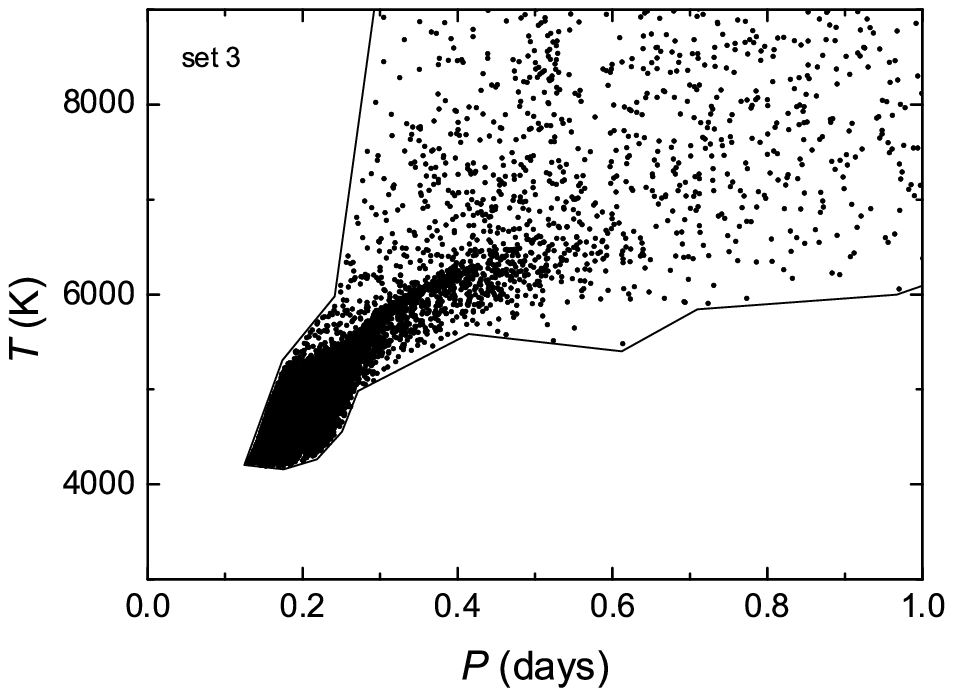,width=7.5cm}} \caption{The
distribution of theoretical contact binaries (set 3) in the $P-T$
plane. Solid curves mark the parameter region of the distribution of
contact binaries at the moment of their formation.} \label{fig3}
\end{figure}

\begin{figure}
\centerline{\psfig{figure=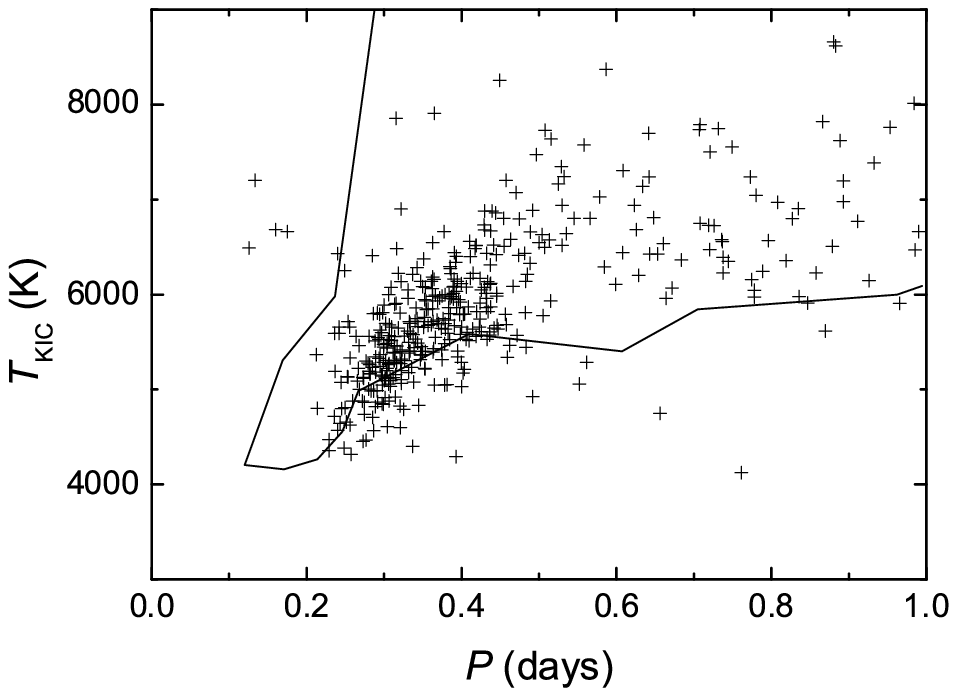,width=7.5cm}} \caption{The
distribution in the $P-T_{\rm KIC}$ plane for observed contact
binaries in the Eclipsing Binary Catalogue given by \citet{Prsa
2011}, \citet{Slawson 2011} and \citet{Matijevic 2012}. Solid curves
are the region of the distribution of contact binaries at the moment
of their formation given in Fig. 7.} \label{fig3}
\end{figure}

We compare the distribution of theoretical contact binaries with
that of the observed contact binaries in orbital period-temperature
plane. Fig. 7 shows the theoretical distribution of contact binaries
at the moment of their formation, and the temperature of contact
binaries ($T$) is obtained according to
$T^4=(R_1^2T_1^4+R_2^2T_2^4)/(R_1^2+R_2^2)$, where $R_1, R_2, T_1$
and $T_2$ are the radii and the effective temperatures of the
primary and the secondary at the moment of the formation of contact
binaries. Solid curves mark the region of the distribution of
contact binaries at the moment of their formation. For clarity, we
only show the distribution for the simulation of set 3, as the other
simulations give similar results. In Fig. 8, we present the
distribution of observed contact binaries in the Eclipsing Binary
Catalogue given by \citet{Prsa 2011}, \citet{Slawson 2011} and
\citet{Matijevic 2012}. Solid curves are the region of the
theoretical distribution of contact binaries given by Fig. 7. It is
obvious that the distribution of the theoretical contact binaries is
in agreement with the observations, although the observed contact
binaries in the highest-density region have longer orbital period,
higher temperature than the theoretical contact binaries. This
difference will be discussed in Section 4.

It is noted that there is no observed contact binaries with $T_{\rm
KIC}<4000$\,K in Fig. 8, although there are some detached binaries
with $T_{\rm KIC}<4000$\,K shown in Fig. 4 and about 59 candidates
of detached binaries with two M dwarfs found by \citet{Becker 2011}.
These detached binaries with very low mass would experience unstable
mass transfer and could not form stable contact binaries
\citep{Jiang 2012}. In Fig. 8, some observed systems are far away
from the theoretical region, in the upper-left corner ($P<0.2$\,d
and $T_{\rm KIC}>6000$\,K) and the lower-right corner ($P>0.5$\,d
and $T_{\rm KIC}<5000$\,K), and they might not be contact binaries.
For the systems in the upper-left corner, the high temperature is
hard to reconcile with the binaries with $P<0.2$\,d that should have
M-dwarf components. They might be the class of ellipsoidal variable
that exhibit sinusoidal variations \citep{Prsa 2011}. In the
lower-right corner, the evolutionary effects can not produce such
low temperature systems from systems with $T_{\rm KIC}>6000$\,K, or
produce such long period systems from systems with $P<0.4$\,d
\citep{Rubenstein 2001}. These systems in the lower-right corner
might have aliased periods \citep{Rucinski 1998}.

\subsection{The birth rate of contact binaries}

%\subsection{The distribution of contact timescale}
\begin{figure}
\centerline{\psfig{figure=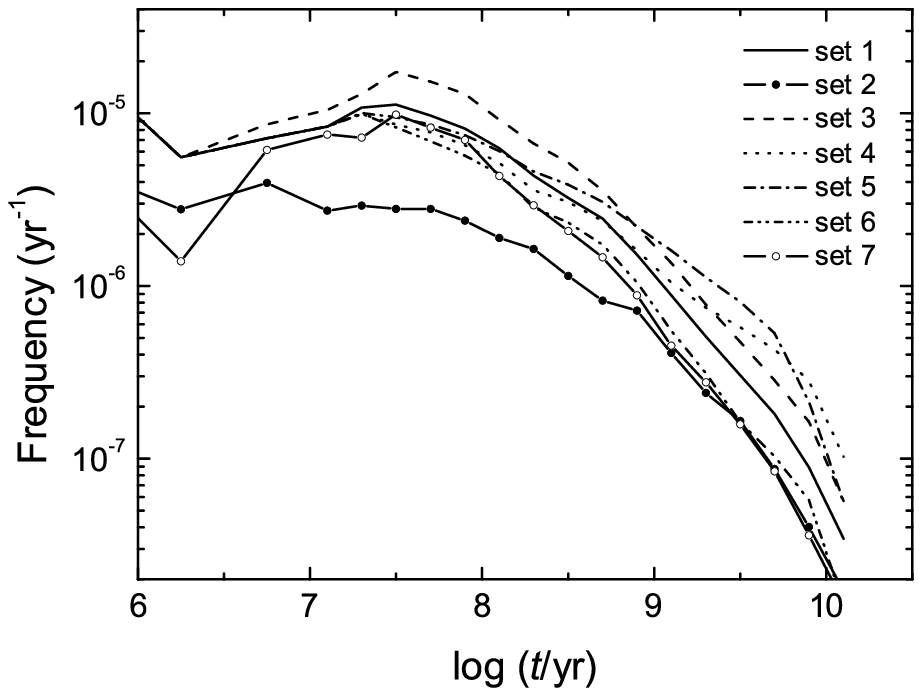,width=7.5cm}} \caption{Evolution
of the birth rates of contact binaries for a single starburst of
$10^6$ in each simulation sets.} \label{fig3}
\end{figure}

\begin{figure}
\centerline{\psfig{figure=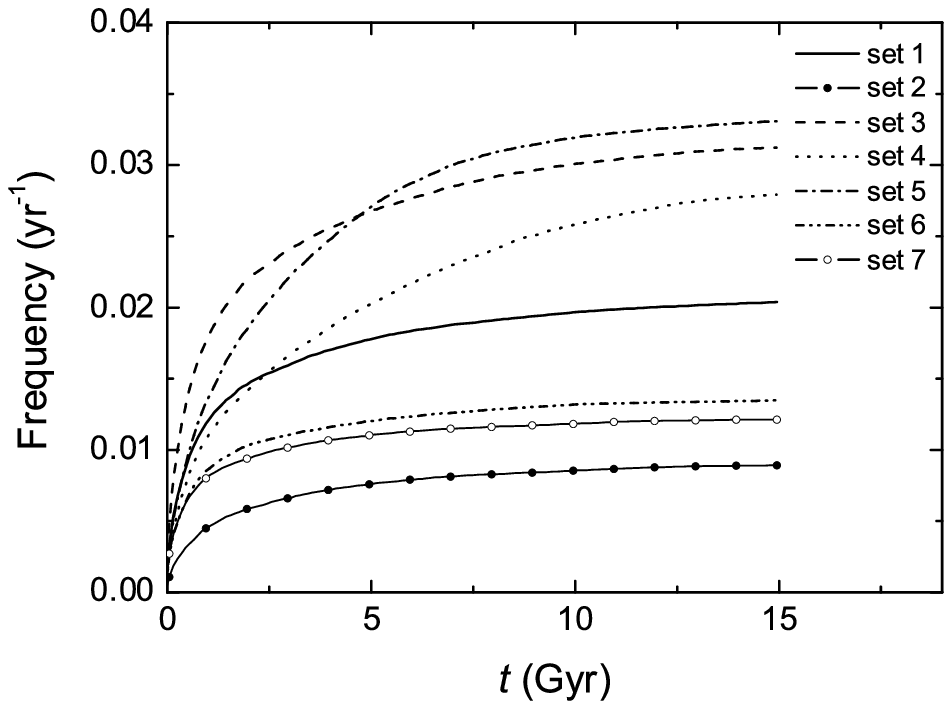,width=7.5cm}} \caption{Similar to
Fig. 9, but for a constant SFR (5\,M$_{\rm \odot}$yr$^{-1}$).}
\label{fig3}
\end{figure}

\begin{figure*}
\centerline{\psfig{figure=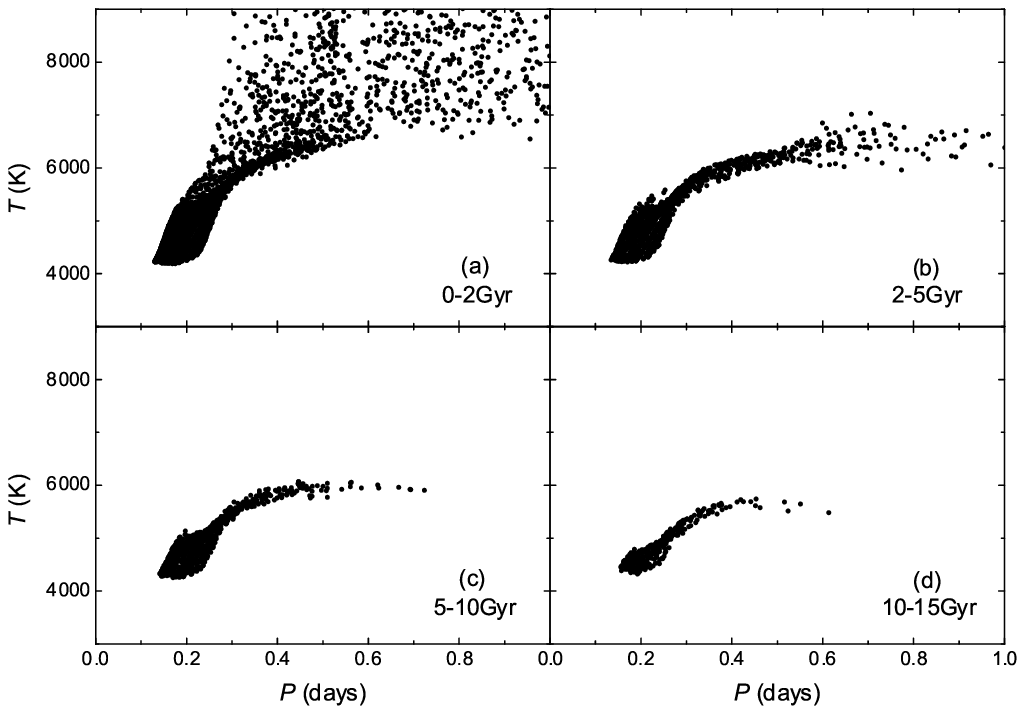,width=15cm}} \caption{Similar to
Fig 7, where contact binaries are divided into four group according
to the age of their formation: (a) 0-2\,Gyr; (b) 2-5\,Gyr; (c)
5-10\,Gyr; (d) 10-15\,Gyr. } \label{fig2}
\end{figure*}

Fig. 9 displays the evolution of the birth rates of contact binaries
for a single starburst of $10^6$ in each simulation sets. In this
figure, we see that the birth rate of contact binaries is in the
range from $1.7\times10^{-8}$\,yr$^{-1}$ to
$1.7\times10^{-5}$\,yr$^{-1}$, and the formation timescale of
contact binaries has a large range from $\sim$1\,Myr to 15\,Gyr. The
birth rates are $0.1-1.7\times10^{-5}$\,yr$^{-1}$ at $1-30$\,Myr.
For age older than about 30\,Myr, the birth rate decreases with
increasing age. Fig. 10 shows birth rates of contact binary for a
constant SFR. The simulations in set 3-5 give a birth rate of
$2.5-3.1\times10^{-2}$\,yr$^{-1}$ for a population older than 10Gyr.
However, the birth rates are lower in the simulations with different
expression for the dependence of MB on $q_{\rm conv}$ (set 1),
different initial mass-ratio distribution (set 2), different initial
period distribution (set 7) or no MB (set 6).

%\subsection{The timescale of the formation of contact binaries from detached binaries}

In order to investigate the characteristics of young contact
binaries and old contact binaries, we show the $P-T$ distribution of
contact binaries at various ages of their formation for a single
starburst in Fig. 11. It shows that the region of contact binaries
decreases significantly with increasing age. The young contact
binaries at an age of $0-2$\,Gyr can occur in the region of
temperature from 4000 to 9000\,K, while old contact binaries at
$10-15$\,Gyr have a temperature lower than 6000\,K. Moreover, the
upper limit of orbital period decreases from $>1$\,d at $0-2$\,Gyr
to 0.65\,d at $10-15$\,Gyr.

\section{discussion and conclusion}
In this paper, we investigated the detached-binary channel for the
formation of contact binaries by carrying out a detailed binary
population synthesis study. We obtained the parameter region for
detached binaries that evolve into contact and the distribution of
contact binaries at the moment of their formation.

The formation of contact binaries in the detached-binary channel
depends on many uncertain input parameters. The main uncertainty
lies in the evolution of orbital period affected by MB. We found
that the dependence of MB on $q_{\rm conv}$ have a significant
influence on the evolution of orbital period of detached binaries,
and therefore on the binary parameter space that produces contact
binaries, the prominent peak in the distribution of orbital periods
and the birth rate of contact binaries. In addition, we varied the
initial mass-ratio distribution to investigate the dependence of the
formation of contact binaries on this model parameter. The
mass-ratio distribution for uncorrelated component masses (set 2) is
more likely to have a very low mass secondary, as compared to the
constant initial mass-ratio distribution (set 1). This leads to
dynamically unstable mass transfer in most cases \citep{Han 2003}.
As a result, the birth rate of contact binaries is greatly reduced
as shown in Fig. 9 and Fig. 10. The initial orbital period
distribution is another important parameter, and it affects very
much the distribution of orbital period of contact binaries and the
birth rate of contact binaries. This is because fewer detached
binaries located in the region of the formation of contact binary in
set 7 lead to fewer contact binaries formed.

\citet{Paczynski 2006} considered detached binaries with $P<1$\,d
might be the progenitors of contact binaries, and found that the
number of these observed detached binaries appear to be insufficient
to produce the number of observed contact binaries based on ASAS
data. By considering the effect of MB, we found that for detached
binaries that evolve into contact, the upper limit of the initial
orbital period could reach about $3-4.2$\,d. Our results agree with
the suggestion given by \citet{Vilhu 1982} that the typical
progenitors of contact binaries are detached binaries with periods
initially shorter than 4\,d. Combining the Eclipsing Binary
Catalogue \citep{Prsa 2011, Slawson 2011, Matijevic 2012}, we found
that the ratio of the birth rate of the progenitors of contact
binaries to that of contact binaries is greater than about 1.2.
Therefore, for the detached-binary channel, the progenitors can be
sufficient to produce observed contact binaries.

\citet{Slawson 2011} found that the period distribution of contact
binaries in the Eclipsing Binary Catalogue has a prominent peak and
a broader component. We found that the period distribution of
contact binaries formed in the detached-binary channel has a peak
and a long tail extending beyond 1\,d, which is very similar to that
of the observed contact binaries as shown in Fig. 6. Contact
binaries in the peak are mainly formed by the effect of MB, and this
prominent peak results from the decrease of AML rate with increasing
orbital period and the short-period limit produced by unstable mass
transfer. For the tail beyond 0.5\,d, almost all contact binaries
are produced by evolutionary expansion of the components. Therefore,
the detached-binary channel can explain the shape of the period
distribution of the observed contact binaries with a peak and a long
tail.

We found a translation of observed contact binaries to longer period
relative to the theoretical contact binaries in the distribution of
orbital period, and a difference in $P-T$ diagram that the observed
contact binaries in the highest-density region have longer orbital
period, higher temperature than the theoretical contact binaries. A
partial explanation might be that low-mass stars have larger radii
than assumed in the models. Observations show that the
sub-solar-mass components of detached binaries have significantly
larger radii than single-star models \citep{Torres 2010}. Another
reason might be that we compute the theoretical distribution of
contact binaries at the moment of their formation. The components
just fill their Roche lobes and are not in good thermal contact. The
period of contact binaries is expected to increase when two
components reach good thermal contact \citep{Li 2004a, Li 2004b, Li
2005}. More importantly, in the subsequent evolution of contact
binaries, the mass ratio becomes smaller and the mass of the
primaries (the massive components) increase during the cyclic
evolution of thermal relaxation oscillation \citep{Robertson 1977,
Li 2004b}. This leads to an increase of orbital period from 0.29 to
0.37d when the mass ratio decreases from 0.6 to 0.22 as shown by
\citet{Rahunen 1981}. In addition, as the primary mass increases,
the temperature of contact binaries would increase. Hence, the
observed contact binaries in the highest-density region in $P-T$
diagram have longer orbital period, higher temperature than our
model contact binaries.

In the detached-binary channel, the formation timescale of contact
binaries has a large range from $\sim$1\,Myr to $15$\,Gyr for a
single starburst. Therefore, the detached-binary channel can explain
the formation of contact binaries in intermediate-age or old cluster
as suggested by \citet{Rucinski 1998}. Furthermore, this channel can
explain the existence of very young contact binaries, such as a
population of ($<0.5$\,Gyr) contact binaries in Moving Group
(kinematically coherent group of stars that share a common origin)
\citep{Bilir 2005}, and two contact binaries as candidate members 25
Ori or Orion OB1a association with age of $~7-10$\,Myr \citep{van
Eyken 2011}, which are believed too young to be formed by detached
binaries. In addition, it is found that the birth rate of contact
binaries decreases with increasing age for age older than about
30\,Myr. The main reason is that binaries with massive components
have short MS lifetime. With the increase of time, the range of the
primary mass decreases for detached binaries with two MS components
that might evolve into contact. Moreover, the decrease of primary
mass range results in the decrease of the upper limit of temperature
and the upper limit of orbital period as shown in Fig. 11.
Therefore, at the moment of the formation of contact binaries, young
contact binaries could have larger range of period and temperature
than old contact binaries.

In our study, we did not consider the evolution of contact phase,
which has been investigated and discussed by many authors, e.g.
\citet{Webbink 1976a, Webbink 1976b}, \citet{Kahler 2002a, Kahler
2002b, Kahler 2004}, \citet{Li 2004a, Li 2004b, Li 2005} and
\citet{Yakut 2005}. The evolution of contact binaries should be
considered in the further BPS study of the detached-binary channel.
In addition, the effect of the third body might be also important
and should be considered because it could make more binaries with
longer period evolve into the formation region of contact binaries
in the detached-binary channel.

\section*{ACKNOWLEDGEMENTS}
It is a pleasure to thank an anonymous referee for his/her many
suggestions and comments which considerably improved the paper. We
thank Dr. Andrej Pr\v{s}a for his great help. This work was
supported by the Chinese Natural Science Foundation (11073049,
11033008, 11103073 and 11373063) and the Western Light Youth
Project.

\bsp

\label{lastpage}

\end{document}